\documentstyle[11pt,newpasp,twoside,epsf]{article}
\def\edcomment#1{\iffalse\marginpar{\raggedright\sl#1\/}\else\relax\fi}
\reversemarginpar

\begin{document}
\title{Maser and Thermal Methanol Emission in the Millimeter Wave Range:
New Masers at 1.3~mm and 2.8~mm
}
\author{S.V. Kalenskii, V.I. Slysh, and I.E. Val'tts}
\affil{Astro Space Center, Lebedev's Physical Institute,
Profsoyuznaya str. 84/32, 117810 Moscow, Russia}

\begin{abstract}
Results of a survey of Galactic star-forming regions in the lines of methanol
$8_{-1}-7_0E$ at 229.8~GHz, $3_{-2}-4_{-1}E$ at 230.0~GHz,
$0_0-1_{-1}E$ at~108.9 GHz, and a series of methanol lines $J_1-J_0E$ 
near 165~GHz are presented. Two masers, DR~21(OH) and DR~21~West,
and two maser candidates,  L~379~IRS3 and NGC 6334I(N),
as well as 16~thermal sources are found at 229.8~GHz. This is the first
detection of methanol masers at a wavelength as short as 1~mm.
At 108.9~GHz, masers were found towards G345.01+1.79 and probably, towards
M~8E. Thermal emission is found towards 28 objects. Only thermal
emission was found at 165 and 230.0~GHz (20 and 7 sources, respectively).
The masers at 229.8~GHz belong to class I, whereas those at 108.9~GHz belong 
to class II, according to the classification by Menten (1991). 
The masers in
DR~21(OH) and DR~21~West can be roughly 
fitted by models with the gas kinetic temperature of the order 
of 50~K. The detection of the 108.9~GHz masers towards G345.01+1.79 and M~8E
may indicate on a specific geometry of these objects. 
The combination of the existence of the class~II $J_0-J_{-1}E$
masers towards W~3(OH), G345.01+1.79, W~48, and Cep~A and our non-detection
of the $3_{-2}-4_{-1}E$ and $J_1-J_0E$ lines is an evidence that
the class~II masers in these objects are pumped by the radiation of hot dust
rather than by that of UC HII-regions.
\end{abstract}

\section{Introduction}

Methanol masers were found in hundreds of Galactic star-forming regions.
According to the classification by Menten (1991), all methanol masers can
be divided into two classes, I and II. The strongest class I masers were
found in the $7_0-6_1A^+$ and the $4_{-1}-3_0E$ transitions at 7 and 8~mm,
respectively. The weaker $5_{-1}-4_0E$, $6_{-1}-5_0E$, and $8_0-7_1A^+$
masers emit at shorter wavelengths, 3 and 2~mm. The strongest class II masers,
$2_{-1}-3_0E$ and $5_1-6_0A^+$ are observed
in the centimeter wave range, and the weaker $3_0-4_1A^+$ and
$J_0-J_{-1}E$ masers are observed at 3 and 2 mm.
Theoretical models of methanol excitation demonstrate that
the class I masers are pumped by collisions,
whereas the class II masers are pumped by external
radiation (Sobolev~et~al.~1997). As a rule,
methanol maser sources emit in several lines of the same class.

The dependence of maser intensity on wavelength, noted above, appears
since both the degree of inversion and
line opacities diminish with the decrease of wavelength. To determine
the parameters of maser sources one must know the shortest wavelength
at which the masers exist. Therefore we performed a search for
the $8_{-1}-7_0E$ masers at 229.8~GHz (1 mm).
No searches for methanol masers have been performed at so high frequencies
before.
The $8_{-1}-7_0E$ transition belongs to the class I, and hence the masers
at 229.8 GHz can be observed together with other masers of this class.
The observations were carried out with the 30-m IRAM telescope.
Besides the $8_{-1}-7_0E$ line, we observed 
the $0_0-1_{-1}E$ line at 108.9~ççà (3~mm), the series of the $J_1-J_0E$ 
lines at 165~GHz (2 mm), and the $3_{-2}-4_{-1}E$ line at 230.0 GHz (1~mm).

\begin{figure}
\plotfiddle{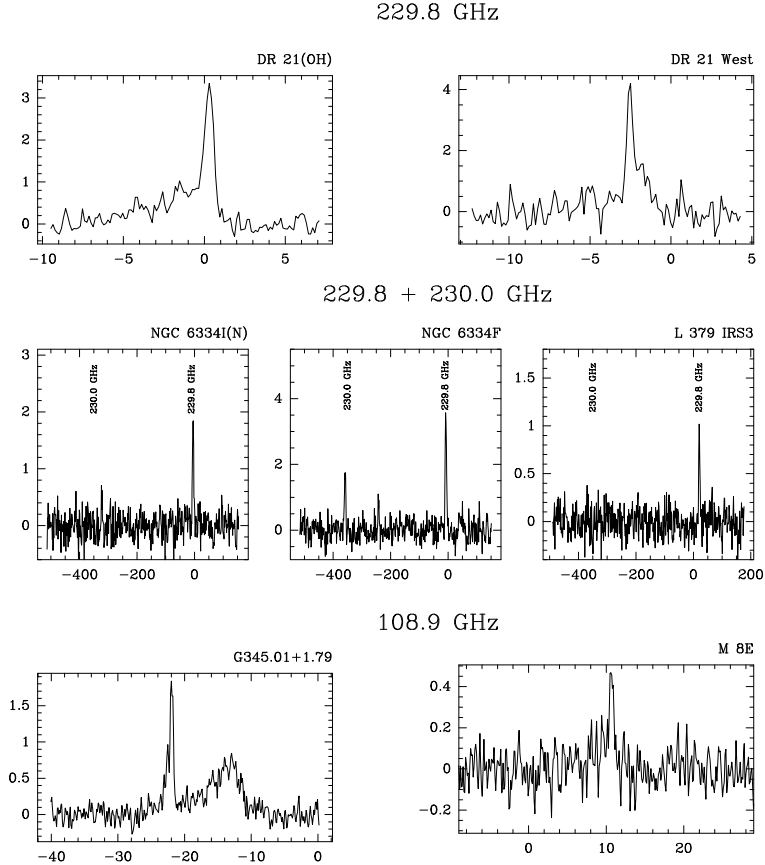}{110mm}{0}{60}{60}{-150}{-135}
\caption{Spectra of maser and thermal emission at 229.8 and 108.9~GHz.
x-axis, $V_{LSR}$; y-axis, $T_R^*$. 1K of $T_R^*$ corresponds
to 10.5~Jy at 229~GHz, and to 6.3~Jy at 108~GHz. 
DR~21(OH), DR~21 West, NGC~6334I(N), and L~379~IRS3 are class I masers,
G345.01+1.79 and M~8E are class II masers. NGC~6334F is a thermal source.
\label{sp1}}
\end{figure}

\section{Results}

In the $8_{-1}-7_0E$ transition narrow maser features were found towards 
DR~21(OH) and DR~21~West (Fig.~1). Thus, for the first time methanol masers
at 1~mm were found. Broad quasithermal lines were detected in 18 sources.
We believe that at least in two of them, L~379IRS3 and NGC~6334I(N) these lines
consist of blended narrow maser features, since other class I lines, detected
in these sources are blends of several maser components
(Slysh~et~al.,~1997; Kogan and Slysh,~1998). 

In the $0_0-1_{-1}E$ line, maser emission was found towards G345.01+1.79
and probably, towards M~8E. In addition, 28 thermal sources were found
in this line, including marginally detected galaxy IC~342. Only thermal
lines were found in the $3_{-2}-4_{-1}E$ and $J_1-J_0E$ transitions.

\section{Class I masers}

\begin{table}[h]
\caption{The observed and modelled flux densities in the class~I lines.
The model parameters:
DR~21(OH), $T_{\rm kin}=60$~K, $n_{\rm H_2}=5.6\times 10^4$~cm$^{-3}$, 
$n_{\rm H_2}/(dV/dR)=6.7\times 10^{-3}$~cm$^{-3}$/(km/s~pc$^{-1})$.
DR~21~West,
$T_{\rm kin}=55$~ë, $n_{\rm H_2}=3.2\times 10^4$~cm$^{-3}$, 
$n_{\rm H_2}/(dV/dR)=6.7\times 10^{-3}$~cm$^{-3}$/(km/s~pc$^{-1})$.
\label{chi}}
\begin{center}
\medskip
\begin{tabular}{|l|rrrr|}
\hline\noalign{\smallskip}
                & \multicolumn{4}{c}{Flux density, Jy}  \\
                & \multicolumn{2}{c}{DR~21(OH)}  
                               &  \multicolumn{2}{c}{DR~21~West}\\
Transition      & model & obs. & model & obs    \\
\noalign{\smallskip}
\hline\noalign{\smallskip}
$4_{-1}-3_0E$   & 15.0  & 15    & 50.0  & 50    \\
$4_0-3_1E$      &  0.4  &$<$0.7 &  1.2  &$<$2.8 \\
$5_{-1}-4_0E$   & 37.8  & 135   & 82.0  & 100   \\
$6_{-1}-5_0E$   & 45.6  &67     & 82.5  &  64   \\
$8_{-1}-7_0E$   & 32.7  &25     & 40.0  &  34   \\
$9_{-1}-8_{-2}E$& 0.003 &$<$0.23& 0.004 &$<$0.25\\
$            $  &       &       &       &       \\
$7_0-6_1A^+$    & 35.9  &340    & 52.5  & 240   \\
$8_0-7_1A^+$    & 27.2  &175    & 28.5  & 150   \\
\noalign{\smallskip}
\hline\noalign{\smallskip}
\end{tabular}
\end{center}
\end{table}

To determine
the conditions, which are necessary to excite the $8_{-1}-7_0E$ masers,
we calculated a number of LVG models with gas kinetic
temperature 10--100~K,
density $10^4$--$10^8$ cm$^{-3}$, and
methanol density divided by velocity gradient  
$7\times 10^{-6}$--$7\times 10^{-2}$~cm$^{-3}$~(km/s~pc$^{-1}$)$^{-1}$.
Any external radiation, except the microwave background, was set equal to  
zero.                                         

The modeling showed that the $8_{-1}-7_0E$ masers having intensities
comparable to those of other class I
masers can arise at gas temperature 30--50~K, typical for the Galactic
molecular clouds. Table~1 demonstrates the observed
flux densities in various class I lines for DR~21(OH) and DR~21~West
as well as flux densities in the best-fit models which were found using
the $\chi^2$ criterion. None of the models represents well the observational
data. This fact is probably related to the model
simplifications, e.g., to the assumption that the same regions emit in
all lines. Actually, maser regions may consist of several layers, 
and different maser lines may form in different layers.

\section{Class II masers}
Class II maser emission was found only in the $0_0-1_{-1}E$ transition.
Masers were detected towards G345.01+1.79, and possibly, towards M~8E.

The class II masers arise owing to a strong external radiation, which
can be produced either by hot dust or by free-free transitions of electrons
in HII-regions. We modelled the methanol excitation under the influence of
the radio emission of HII-regions. A number of models accounting for the
hot dust emission was calculated by Sobolev~et~al.~(1997).

Slysh~et~al. (1995) detected the $J_0-J_{-1}E$ masers towards
W~3(OH), G345.01+1.79, W~48, and Cep~A. Our modelling of the class II masers
pumping by the radio emission of HII-regions showed that this pumping
always produce the $3_{-2}-4_{-1}E$, $J_1-J_0E$, and the $5_{-2}-6_{-1}E$
masers, comparable in intensity with the $J_0-J_{-1}E$ masers. However,
the $3_{-2}-4_{-1}E$ and $J_1-J_0E$ masers were not found in this survey,
and the $5_{-2}-6_{-1}E$ masers were not found by Slysh et al.~(1997).
At the same time, the models of methanol pumping by the emission of hot dust,
published by Sobolev~et~al.~(1997) predict that the intensity of the
$J_0-J_{-1}E$ masers is larger than that of the $J_1-J_0E$ masers,
and no strong maser emission arise in the $3_{-2}-4_{-1}E$ and
$5_{-2}-6_{-1}E$ lines, in a qualititative agreement with the observations.

Thus, the results of our observations, combined with the results by
Slysh et al.~(1995, 1997) mean that the hot dust
emission must be taken into account in the modelling of class II methanol
masers. 

Our modelling shows that the $0_0-1_{-1}E$ transition
is thermalized with the
increase of either density or methanol content much earlier than any other
class II line. 
This fact explains why the masers at 108.9~GHz are so rare. The masers
in G345.01+1.79 and M~8E are likely to be either the least dense class II
masers, or their geometry (e.g., disk-like structure) provides an efficient
photon escape.



\acknowledgements The authors are grateful to the Pico Veleta observatory
staff for the help during the observations.
The work was done under a partial financial support from the International
Scientific Foundation (grant No. MND300) and the Russian Foundation for
Basic Research (grants No. 95-02-05826 and 98-02-16916).

\end{document}